\documentclass[twocolumn,showpacs,showkeys]{revtex4}
\usepackage{graphicx,subfigure}

\newcommand{\bastar}{\begin{eqnarray*}}
\newcommand{\eastar}{\end{eqnarray*}}
\newskip\humongous \humongous=0pt plus 1000pt minus 1000pt

\newif\ifdtup

\relax
\newcommand{\bea}{\begin{eqnarray}}
\newcommand{\eea}{\end{eqnarray}}
\newcommand{\nn}{\nonumber}
\newcommand{\pro}{\partial}
\newcommand{\oneg}{\dfrac{1}{g}}
\newcommand{\dfrac}{\displaystyle\frac}
\newcommand{\abc}{\alpha \beta \gamma}
\newcommand{\e}{\vec e}

\newcommand{\m}{\vec m}

\newcommand{\vl}{{\bf {l}}}
\newcommand{\vk}{{\bf {k}}}
\newcommand{\vp}{{\bf {p}}}
\newcommand{\X}{{\vec X}}

\newcommand{\A}{{\vec A}}
\newcommand{\B}{{\vec B}}

\newcommand{\F}{{\vec F}}

\newcommand{\vGm}{{\bf \Gamma}}

\newcommand{\vR}{{\bf R}}

\newcommand{\vOm}{{\bf \Omega}}

\newcommand{\vtl}{\tilde{{\bf l}}}

\newcommand{\vtp}{\tilde{{\bf p}}}

\newcommand{\hn}{{\hat n}}

\newcommand{\hA}{{\hat A}}

\newcommand{\mn}{{\mu\nu}}

\begin{document}
\title {Topology of Vacuum Space-Time}
\author{Y. M. Cho}
\email{ymcho@yongmin.snu.ac.kr}
\affiliation{Center for Theoretical Physics
and School of Physics, College of Natural Sciences,
Seoul National University,
Seoul 151-742, Korea  \\}
\begin{abstract}
~~~~~We present a topological classification of vacuum space-time.
Assuming the $3$-dimensional space allows a global chart, we show that
the static vacuum space-time of Einstein's theory
can be classified by the knot topology $\pi_3(S^3)=\pi_3(S^2)$.
Viewing Einstein's theory as a gauge theory of
Lorentz group and identifying the gravitational connection
as the gauge potential of Lorentz group, we construct all possible
vacuum gravitational connections which give a vanishing
curvature tensor. With this we show that
the vacuum connection has the knot topology,
the same topology which describes the multiple vacua of $SU(2)$
gauge theory. We discuss the physical implications of our result
in quantum gravity.
\end{abstract}
\maketitle

\section{Introduction}

Gauge theories and general relativity have many things in common.
Both have highly non-linear self-interaction.
This is not accidental. Actually they are based on the same
principle, the general invariance or equivalently the gauge invariance.
The gauge theory is well-known to be a part of
a higher-dimensional gravity which originates from the
extrinsic crvature of a non-trivial
embedding of the $4$-dimensional space-time to the
$(4+n)$-dimensional unified space \cite{kal,jmp75,prd75}.
When the unified space has an $n$-dimensional isometry,
the $(4+n)$-dimensional Einstein-Hilbert action
describes the gauge theory of the isometry group $G$ coupled to
the $4$-dimensional Einstein's theory \cite{jmp75,prd75}.
Conversely, Einstein's theory itself can be viewed as
a gauge theory, because the general invariance of
Einstein's theory can be viewed as a gauge
invariance \cite{kib,uti,prd76a,prd76b}.
It can be viewed either as a gauge invariance of
$4$-dimensional translation group \cite{kib,prd76a},
or as a gauge invariance of Lorentz group \cite{uti,prd76b}.

On the other hand, it has been well-known that the
non-Abelian gauge theory has a non-trivial topology.
For example, the non-Abelian gauge theory admits
magnetic monopoles which can be classified by the
monopole topology $\pi_2(S^2)$ \cite{prd80,prl80}.
Moreover, the non-Abelian gauge theory has a mutiple
vacua which can be classified by the knot topology
$\pi_3(S^3)=\pi_3(S^2)$ \cite{bpst,thooft,plb79,baal,plb06}.
If so, one might suspect that Einstein's theory
should also allow simliar non-trivial topology.
This turns out to be the case. {\it The purpose of this Letter is
to demonstrate that Einstein's theory has
exactly the same multiple vacua as the $SU(2)$ gauge theory.
We show that, assuming the $3$-dimensional space admits a global chart,
the topology of the static vacuum space-time can be classified
by the knot topology $\pi_3(S^3)=\pi_3(S^2)$.}
%This tells that in $R^4$ space-time
%the vacuum gravitational connection
%of Einstein's theory which produces a vanishing curvature
%tensor has mathematically identical topological structure
%we have in the vacuum potential of $SU(2)$
%gauge theory.

As we have remarked, Einstein's theory can be viewed as
a gauge theory \cite{kib,uti}. There are two ways to view
Einstein's theory as a gauge theory \cite{prd76a,prd76b}.
It can be viewed as a gauge theory of the $4$-dimensional
translation group \cite{prd76a}. In this view the (non-trivial part of)
the tetrad becomes the gauge potential of translation group.
Or else it can be viewed
as a gauge theory of Lorentz group \cite{prd76b}.
In this view the gravitational connection (more precisely
the spin connection) becomes the gauge potential of Lorentz group
and the curvature tensor becomes the gauge field strength.
Here we adopt the second view, and construct
all possible vacuum gravitational connection
which yields a vanishing curvature tensor (field strength).
With this we show that the vacuum gravitational connection
is described by the vacuum potential of the $SU(2)$ subgroup
of Lorentz group, which confirms that the static vacuum space-time
has the knot topology.

To understand the topology of vacuum space-time it is
crucial to understand the topology of non-Abelian gauge theory.
A best way to describe the topology of gauge theory is
through the magnetic symmetry \cite{prd80,prl81}.
In non-Abelian gauge theory it is well-known that
one can restrict and/or decompose the gauge potential imposing
a magnetic symmetry.
Consider the $SU(2)$ gauge theory for simplicity, and let $\hn$ be
an arbitrary unit isotriplet. We can restrict the gauge potential
imposing the following magnetic symmetry which dictates
$\hn$ to be invariant under the parallel transport \cite{prd80,prl81},
\bea
D_\mu \hn=(\pro_\mu +g \vec A_\mu \times)~\hn=0.~~~~~(\hn^2=1)
\label{ccon}
\eea
This selects the restricted potential $\hA_\mu$,
\bea
&\hA_\mu= A_\mu \hn- \dfrac 1g \hn \times \pro_\mu \hn,
~~~(A_\mu= \hn \cdot \A_\mu)
\label{rp}
\eea
as the potential which satisfies (\ref{ccon}). Such a condition
is called a magnetic symmetry because this automatically projects
out the potential which describes the non-Abelian monopole.
Indeed with $A_\mu=0$ and $\hn=\hat r$, (\ref{rp}) describes
the well-known Wu-Yang monopole \cite{prd80,prl80}.
In this case the monopole topology $\pi_2(S^2)$ is described by
$\hn$ which defines the mapping from $S^2$ of $3$-dimensional
space surrounding the monopole core to the coset space
$SU(2)/U(1)$ of $SU(2)$.

In gauge theory this way of projecting out
the restricted potential of the maximal Abelian subgroup $H$
of the gauge group $G$ is called
the Abelian projection \cite{prd80,prl81}.
With the Abelian projection we can retrieve the full gauge
potential simply by adding the gauge covariant valence
potential $\X_\mu$ of the coset space $G/H$ to $\hA_\mu$,
\bea
\A_\mu=\hA_\mu+\X_\mu.
\eea
This is called the Abelian decomposition.
A nice feature of the Abelian decomposition is that it is
gauge independent. Once the Abelian direction $\hn$ is chosen,
the decomposition follows automatically,
independent of the choice of a gauge \cite{prd80,prl81}.
As importantly the Abelian part $\hA_\mu$ retains all topological
properties of the gauge theory, because it has the full
gauge degrees of freedom of $G$. The Abelian decomposition
has played a crucial role for us to understand
the non-Abelian dynamics, in particular the confinement
mechanism in QCD \cite{prd02,jhep04,fadd,kondo}.

A remarkable feature of the magnetic symmetry is that it
forms an isometry \cite{jmp75,prl80}, so that if we have
two magnetic symmetries
\bea
&D_\mu \hn_1=0,~~~D_\mu \hn_2=0,
\eea
we automatically have a third one $\hn_1 \times \hn_2$,
\bea
D_\mu (\hn_1 \times \hn_2)=0.
\eea
Let $\hat n_i~(i=1,2,3)$ be orthonormal isotriplets
which form a right-handed basis $(\hn_1 \times \hn_2=\hn_3)$,
and let
\bea
\forall_i~~~D_\mu \hn_i =0.
\label{vcon}
\eea
Obviously this puts a strong restriction to the gauge potential
and corresponding field strength. Indeed (\ref{vcon}) requires
the potential to have a vanishing field strength.
This is because we have
\bea
\forall_i~~~[D_\mu,~D_\nu]~\hn_i=g \F_\mn \times \hn_i=0,
\eea
which requires $\F_\mn=0$. This tells that
a vacuum potential must be the one
which parallelizes the local orthonormal frame.

Solving (\ref{vcon}) we obtain a most general $SU(2)$
vacuum potential
\bea
&\A_\mu=\hat \Omega_\mu = - C_\mu \hn - \dfrac{1}{g} \hn \times \pro_\mu \hn
= - C_\mu^k~\hn_k, \nn\\
&\dfrac{1}{g} \hn \times \pro_\mu \hn
= C_\mu^1~\hn_1 + C_\mu^2~\hn_2, \nn\\
&C_\mu^k = -\dfrac{1}{2g} \epsilon_{ij}^{~~k} (\hn_i \cdot \pro_\mu \hn_j),
\label{vac}
\eea
where $\hn=\hn_3$ and $C_\mu=C_\mu^3$.
One can easily check that $\hat \Omega_\mu$ describes a vacuum
\bea
&\hat \Omega_{\mu\nu} = \pro_\mu \hat \Omega_\nu
-\pro_\nu \hat \Omega_\mu + g \hat \Omega_\mu \times \hat \Omega_\nu \nn\\
&=-(\pro_\mu C_\nu^k -\pro_\nu C_\mu^k
+ g \epsilon_{ij}^{~~k} C_\mu^i C_\nu^j)~\hn_k = 0.
\label{vacf}
\eea
This tells that $\hat \Omega_\mu$ or equivalently
$(C_\mu^1,C_\mu^2,C_\mu^3)$ describe a classical $SU(2)$ vacuum.
Notice that, although the vacuum is fixed by three isometries,
it is essentially fixed by $\hn$. This is because $\hn_1$
and $\hn_2$ are uniquely determined by
$\hn$, up to a $U(1)$ gauge transformation which leaves $\hn$
invariant. With
\bea
&\hn = \Bigg(\matrix{\sin{\alpha}\cos{\beta} \cr
\sin{\alpha}\sin{\beta} \cr \cos{\alpha}}\Bigg),
\label{n}
\eea
we have
\bea
&C_\mu^1= \oneg (\sin \gamma \pro_\mu \alpha
-\sin{\alpha} \cos \gamma \pro_\mu \beta), \nn\\
&C_\mu^2 = \oneg (\cos \gamma \pro_\mu \alpha
+\sin{\alpha} \sin \gamma \pro_\mu \beta), \nn\\
&C_\mu^3 = \oneg (\cos{\alpha} \pro_\mu \beta+\pro_\mu \gamma),
\eea
where the angle $\gamma$ represents the $U(1)$ angle which leaves
$\hn$ invariant.

A nice feature of (\ref{vac}) is that the topological
character of the vacuum is naturally inscribed in it.
The topological vacuum quantum number is given by
the non-Abelian Chern-Simon index of
the potential $\hat \Omega_\mu$ \cite{thooft,plb79,plb06}
\bea
&Q=-\dfrac{3g^2}{8\pi^2} \int \epsilon_{\abc}
(C_\alpha^i \pro_\beta C_\gamma^i
+\dfrac{g}{3}\epsilon_{ijk} C_\alpha^i C_\beta^j C_\gamma^k) d^3x \nn\\
&=-\dfrac{g^3}{96\pi^2} \int \epsilon_{\abc} \epsilon_{ijk}
C_\alpha^i C_\beta^j C_\gamma^k d^3x,  \nn\\
&(\alpha,\beta,\gamma=1,2,3)
\label{nacsi}
\eea
which represents the non-trivial vacuum topology $\pi_3(S^3)$
of the mapping from the compactified
$3$-dimensional space $S^3$ to the $SU(2)$ space $S^3$.
But this vacuum topology can also be described by $\hn$,
because (with $\hn(\infty)=(0,0,1)$) it defines the mapping $\pi_3(S^2)$
which can be transformed to $\pi_3(S^3)$ through
the Hopf fibering \cite{plb79,plb06}. So both
$\hat \Omega_\mu$ and $\hn$
describe the vacuum topology of the $SU(2)$ gauge theory.
But since $\hat \Omega_\mu$ is essentially fixed by $\hn$ we can
conclude that the vacuum topology is imprinted in $\hn$.

Now, regarding Einstein's theory as a gauge theory of
Lorentz group, we can find the most general vacuum of Einstein's theory
by imposing the magnetic isomstry which gives a vanishing
curvature tensor. Let $J_{ab}=-J_{ba}~(a,b=0,1,2,3)$ be
the six generators of Lorentz group,
\bea
&[J_{ab}, ~J_{cd}] =f_{ab,cd}^{~~~~~~mn}~J_{mn}, \nn\\
&f_{ab,cd}^{~~~~~~mn}=\eta_{ac} \delta_b^{~[m} \delta_d^{~n]}
-\eta_{bc} \delta_a^{~[m} \delta_d^{~n]} \nn\\
&+\eta_{bd} \delta_a^{~[m} \delta_c^{~n]} -\eta_{ad} \delta_b^{~[m}
\delta_c^{~n]},
\label{lgcr}
\eea
where $\eta_{ab}=diag~(-1,1,1,1)$ is the Minkowski metric.
Using the 3-dimensional rotation and boost generators
$L_i$ and $K_i$ we can express the Lorentz algebra as
\bea
& [L_i, ~L_j] = \epsilon_{ijk} L_k,
~~~[L_i, ~K_j] = \epsilon_{ijk} K_k, \nn \\
&[K_i, ~K_j] =-\epsilon_{ijk} L_k.~~~(i,j,k= 1,2,3)
\eea
Now, let $\vp$ (or $p^{ab}$) be an isosextet which forms
an adjoint representation of Lorentz group, and let $\vtp$
(or $\tilde p^{ab}=1/2 \epsilon_{abcd}~p^{cd}$) be its
dual partner. They can be expressed by two isotriplets $\m$ and $\e$,
the magnetic and electric components of $\vp$ which correspond to
the $3$-dimensional rotation and boost,
\bea
\vp=\left( \begin{array}{c} \m \\
\e \end{array} \right),
~~~\vtp=\left( \begin{array}{c} \e \\
-\m \end{array} \right),~~~\tilde {\vtp}=-\vp.
\eea
To proceed further let $\vGm_\mu$ (or $\Gamma_\mu^{~ab}$)
be the gauge potential of Lorentz group which describes the
spin connection, and $\vR_\mn$ (or $R_\mn^{~~ab}$)
be the curvature tensor
\bea
\vR_\mn=\pro_\mu \vGm_\nu-\pro_\nu \vGm_\mu
+\vGm_\mu \times \vGm_\nu,
\eea
where we have normalized the coupling constant to be the unit
which one can always do without loss of generality.
Now, consider the following magnetic isometry
\bea
D_\mu \vp = (\pro_\mu + \vGm_\mu \times) ~\vp=0.
\label{ic}
\eea
This automatically assures
\bea
D_\mu \vtp =(\pro_\mu + \vGm_\mu \times) ~\vtp=0,
\label{dic}
\eea
which tells that the magnetic isometry in Lorentz group
always contains the dual partner. To verify this
we decompose the gauge potential
$\vGm_\mu$ into the 3-dimensional rotation and boost parts
$\A_\mu$ and $\B_\mu$, and let
\bea
\vGm_\mu= \left( \begin{array}{c} \A_\mu \\
\B_\mu \end{array} \right).
\eea
With this both (\ref{ic}) and (\ref{dic}) can be written
as \cite{grg1}
\bea
&D_\mu \m= \B_\mu \times \e,
~~~~D_\mu \e= -\B_\mu \times \m, \nn\\
&D_\mu = \pro_\mu + \A_\mu \times.
\label{ic1}
\eea
This confirms that (\ref{ic}) and (\ref{dic}) are actually identical
to each other, which tells that the magnetic isometry in Lorentz
group must be even-dimensional.

Now, we can project out the vacuum connection which
yields a vanishing curvature tensor with
a proper magnetic symmetry.
Let $\vl_i$ and $\vk_i~(i=1,2,3)$ be the unit sextet vector fields
which describe the rotation and the boost, which form
an orthonormal basis of the adjoint representation of Lorentz group.
And let $\hn_i$ be the orthonormal isotriplets of $SU(2)$ as before.
Identifying $SU(2)$ as a subgroup of Lorentz group,
we have
\bea
&\vl_i= \left( \begin{array}{c} \hn_i \\
0 \end{array} \right),
~~~\vk_i= \left( \begin{array}{c} 0 \\
\hn_i  \end{array} \right)= -\vtl_i.
\label{lbasis}
\eea
Moreover,
\bea
&\vl_i \cdot \vl_j=\delta_{ij},~~~\vl_i \cdot \vk_j=0,
~~~\vk_i \cdot \vk_j=-\delta_{ij},  \nn\\
&\vl_i \times \vl_j= \epsilon_{ijk} \vl_k,
~~~\vl_i \times \vk_j= \epsilon_{ijk} \vk_k, \nn\\
&\vk_i \times \vk_j= -\epsilon_{ijk} \vl_k.
\eea
Now, it must be clear that the magnetic isometry
which selects the vacuum space-time is given by
\bea
&\forall_i~~~D_\mu \vl_i =0,
\label{vic}
\eea
or equivalently
\bea
\forall_i~~~D_\mu \vk_i = -D_\mu \vtl_i= 0.
\label{dvic}
\eea
Actually the vacuum need both. But one is enough because one assures
the other, since $\vk_i=-\vtl_i$.

In $3$-dimensional notation (\ref{vic}) or (\ref{dvic})
can be written as
\bea
&\forall_i~~~D_\mu \hn_i= \B_\mu \times \hn_i,
~~~D_\mu \hn_i= -\B_\mu \times \hn_i.
\label{vic1}
\eea
Clearly this has the solution
\bea
&\A_\mu=\hat \Omega_\mu,~~~\B_\mu=0.
\label{gvac1}
\eea
where $\hat \Omega_\mu$ (with $g=1$) is precisly the vacuum potential
(\ref{vac}) of the $SU(2)$ subgroup. So the vacuum connection $\vOm_\mu$
which yields vanishing curvature tensor is given by
\bea
\vGm_\mu=\vOm_\mu=\left( \begin{array}{c} \hat \Omega_\mu \\
0 \end{array} \right).
\label{gvac2}
\eea
This proves that the gravitational connection of
the vacuum space-time in Einstein's theory is
fixed by the rotational part (or the magnetic
component) of the spin connection which describes
the multiple vacua of $SU(2)$ gauge theory.
This assures that the topology of the static vacuum connection
is identicsl to the topology of the vacuum
potential in $SU(2)$ gauge theory,
which can be classified by the knot topology $\pi_3(S^3)=\pi_3(S^2)$.

Notice that in general (\ref{gvac2}) has meaning only locally
section-wise. So for the topological classification
we must assume that (just as in gauge theory)
the $3$-dimensional space has a global chart.
Another point to keep in mind is that the vacuum space-time
is described in terms of the connection, not the metric.
To obtain the vacuum metric, we must convert the vacuum
connection to the vacuum metric, solving
the first-order equation of the spin connection
expressed in terms of the tetrad.
But we emphasize that the topology of vacuum space-time
becomes transparent only in terms of the connection.
And for the topology of the vacuum we do not need the metric,
which only complicates and obsecures the problem.

We have shown how the magnetic isometry allows us
to construct all possible vacuum cinnections of
Einstein's theory. The magnetic isometry also allows us
to make the Abelian decomposition of Einstein's theory \cite{grg1}.
Imposing the magnetic isometry which selects
the restricted connection of the maximal Abelian subgroup of
Lorentz group, we can have the Abelian decomposition of
Einstein's theory without compromising the general invariance.
In this decomposition Einstein's theory
can be interpreted as a theory of restricted gravity
made of the restricted connection, which has the valence
connection as the gravitational source \cite{grg1}.
Einstein's theory allows two different Abelian decompositions,
the space-like decomposition and the light-like decomposition,
because Lorentz group has two maximal Abelian subgroups.

Clearly our result raises more questions.
A first question is the monopole topology in
Einstein's theory. The above analysis strongly indicates that
the topology of Einstein's theory is closely related to the
topology of $SU(2)$ gauge theory. This implies
that Einstein's theory should also have the $\pi_2(S^2)$
topology which describes a gravito-magnetic
monopole \cite{cho91,new}.
It would be interesting to see how one can embed
the non-Abelian monopole in Einstein's theory to
obtain the gravito-magnetic monopole.

Another question is the vacuum tunneling in quantum gravity.
It has been well-known that the multiple vacua in gauge theory
are unstable against quantum fluctuation,
because they are connected by instantons \cite{bpst,thooft,plb79}.
If so, one may ask whether we can have the gravito-instantons
in Einstein's theory which can connect
topologically distinct vacuum space-times.
The answer is not obvious, because Einstein's theory
and gauge theory have an important difference.
In gauge theory the Yang-Mills action is quadratic in
field strength, but in Einstein's theory the
Einstein-Hilbert action is linear in curvature tensor.
Moreover, in gauge theory the fundamental field which propagates
is the gauge potential but in Einstein's theory the fundamental
field which propagates is the metric
(not the connection) \cite{prd75,prd76b}.
So, even though one can view Einstein's theory as a gauge theory,
the Einstein's eqation is not of Yang-Mills type.
This, of course, does not rule out the gravito-instanton.
There is a real possibility that such
instanton could exist. Here we simply remark that candidates
of the ``gravitational instanton" in Einstein's theory
which has finite Euclidian action have been proposed before,
but without any reference to the multiple
vacua and the quantum tunneling \cite{egu}.

The discussions on these questions and related subjects
will be presented in a separate paper \cite{grg3}.

{\bf ACKNOWLEDGEMENT}

~~~The work is supported in part by the ABRL Program
(Grant R14-2003-012-01002-0) and by the International Coorperation
Program of Korea Science and Enginering Foundation.

\end{document}